\documentclass[11pt]{article}
\usepackage{cite}
\usepackage{amsmath,amsfonts,amssymb}
\usepackage[small,bf,hang]{caption}
\usepackage{graphics}
\usepackage{graphicx}
\DeclareGraphicsExtensions{.pdf}

\def\hybrid{
        \topmargin -20pt
        \oddsidemargin 0pt
        \headheight 0pt \headsep 0pt
        \textwidth 6.25in 
        \textheight 9.5in 
        \marginparwidth .875in
        \parskip 5pt plus 1pt \jot = 1.5ex}

\hybrid

 \csname
@addtoreset\endcsname{equation}{section}

\def\moth{\mathsurround=0pt}
\newdimen\zo \zo=0pt

\def\tick{\leaders\hrule height 0.5ex depth 0pt \hskip 0.5pt}
\def\upboxfill{$\moth \setbox\zo\hbox{\tick}%
  \hskip 3pt\hbox to 0pt{$\tick$\hss}\hrulefill \hbox to 7.5pt{$\tick$\hss}$}

\def\dtick{\leaders\hrule height .34pt depth 0.5ex \hskip 0.5pt}
\def\downboxfill{$\moth \setbox\zo\hbox{\dtick}%
  \hskip 2pt\hbox to 0pt{$\dtick$\hss}\hrulefill \hbox to 2pt{$\dtick$\hss}$}

\def\bec{\begin{center}}
\def\ec{\end{center}}

 \def\det{{\rm det\,}}
\def\be{\begin{equation}}
\def\ee{\end{equation}}
\def\bea{\begin{eqnarray}}
\def\eea{\end{eqnarray}}
\def\ba{\begin{array}}
\def\ea{\end{array}}

\begin{document}

\begin{titlepage}
\begin{center}
\vskip 2.5cm
{\Large \bf {
Holographic zero sound for $\mathcal{N}=2$ SCFTs in 4d}}\\
\vskip 1.0cm
{\large {Dibakar Roychowdhury}}
\vskip 1cm
{\it {Department of Physics}}\\
{\it {Indian Institute of Technology Roorkee}}\\
{\it {Roorkee 247667, Uttarakhand, India}}\\

\vskip 2.5cm
{\bf Abstract}
\end{center}

\vskip 0.1cm

\noindent
\begin{narrower}
We build up the notion of metallic holography for $\mathcal{N}=2$ SCFTs in four dimensions, in the presence of a finite $U(1)$ chemical potential. We compute two point correlation functions and study their properties in the regime of low frequency and low momentum.  The color sector reveals gapless excitations, one of which propagates with a finite phase velocity while the other mode attenuates through scattering. The flavour sector, on the other hand, reveals spectrum that contains quasiparticle excitations which propagate with a definite phase velocity together with modes that propagate with a finite group velocity which quantum attenuates quite similar in spirit as that of Landau's zero sound modes.  We also compute associated AC conductivities, which exhibit different characteristics for different (color and flavour) sectors of $\mathcal{N}=2$ SCFTs.  
\end{narrower}
\end{titlepage}

\newpage

\tableofcontents
\baselineskip=16pt

\section{Introduction and Overview}
Holography beyond the standard Maldacena conjecture \cite{Maldacena:1997re}-\cite{Witten:1998qj} has gained renewed attention in the recent years. These include examples starting with $ AdS_7 $ \cite{Apruzzi:2013yva}-\cite{Nunez:2018ags} and $ AdS_6 $ \cite{DHoker:2016ujz}-\cite{Legramandi:2021uds}, that goes down to $ AdS_5 $ \cite{Sfetsos:2010uq}-\cite{Gaiotto:2009gz} and many others preserving an AdS factor. However, for the purpose of the present paper, we would be particularly interested in this third category of examples, that are associated with an $ AdS_5 $ factor as a solution in type IIA supergravity and has been explored in numerous directions untill recent years \cite{Itsios:2012dc}-\cite{Macpherson:2024frt}. These geometries are categorised under the Gaiotto-Maldacena (GM) class of solutions \cite{Gaiotto:2009gz} that preserve $ \mathcal{N}=2 $ SUSY and are holographically dual to a class of strongly coupled $ \mathcal{N}=2 $ SCFTs \cite{Gaiotto:2009we} in 4d\footnote{This is precisely the low energy (IR) limit where all the NS5 branes are put on top of each other and the world-volume theory is ``effectively'' four dimensional \cite{Lozano:2016kum}-\cite{Nunez:2019gbg} as well as strongly coupled. Here, low energy implies energies much smaller as compared to the inverse size of the D4 world-volume direction ($E \ll x^{-1}_6$) that is extended along the $r$ axis. In other words, the field theory is not excited enough to see this additional world-volume direction ($x_6$) of the D4 brane. The effective gauge coupling on D4 world-volume behaves as $\frac{1}{g^2_{4}}\sim \frac{r_{n+1}-r_n}{g_s \sqrt{\alpha'}}$ \cite{Lozano:2016kum}, where the numerator denotes the separation between two successive NS5 branes located at $ r_{n+1} $ and $ r_n $. The holographic limit corresponds to this strongly coupled fixed point where the separation goes to zero. This is precisely reflected in the $AdS_5$ factor of the Sfetsos-Thompson (ST) solution \eqref{e2.1}, which corresponds to a boundary SCFT living in 4d. Once the background \eqref{e2.1} is realised, the next task is to add probe D-branes that source ``external'' charge carriers to this superconformal quiver living in 4d.}. Some special solutions under the GM class, namely the Abelian T-dual (ATD) and the non-Abelian T-dual (NATD) of $ AdS_5 \times S^5 $ \cite{Lozano:2016kum},  are found to preserve integrability in the Liouvillian sense \cite{Nunez:2018qcj}. 

In our paper, we focus on $ \mathcal{N}=2 $ linear quivers that are dual to NATD of $ AdS_5 \times S^5 $. These SCFTs contain color nodes with a linearly increasing rank and with no flavours included \cite{Lozano:2016kum}. Therefore, these are the quivers that never ``complete'' and are of infinite length. However, the completion of $ \mathcal{N}=2 $ quivers can be achieved by adding flavour D6 branes in the bulk that results into other class of solutions within the GM category which we do not focus here. 

The prime motivation of the present analysis is to unearth $ \mathcal{N}=2 $ SCFTs in the presence of finite density charge carriers carrying a $ U(1) $ charge and explore their two point correlations at low frequency and momentum. These would eventually lead towards various response functions of the theory. One elegant way to achieve this goal is to add probe D-branes that source a charge density ($ \varrho $) and hence a chemical potentials ($ \mu $) for the dual $ \mathcal{N}=2 $ quiver at $ T=0 $.

Probe D-brane captures strange metallic behaviour via holography \cite{Hartnoll:2009ns}-\cite{Karch:2008fa}. In a holographic set up, probe D-brane represents                                                                                                                finite density charge carriers in a strongly interacting medium. The strongly interacting medium acts like a conformal fixed point that is dual to background spacetime in 10d on which the D-brane is propagating. The quantum critical points, that we would be interested in, are those preserving $ \mathcal{N}=2 $ SUSY and are charactersied by the dynamic exponent, $ z=1 $. It turns out that quantum critical points with $ z=1 $, exhibit quasiparticle excitations (at $ T=0 $) that appear as a pole in the charge density and current-current retarded two point function \cite{Mateos:2007vc}-\cite{Edalati:2013tma}, in the domain of low frequency ($ \omega $) and low momentum ($ k $). This is a typical feature that characterises so called ``quantum liquids'' \cite{b1} that are transnationally invariant systems at $ T=0 $ and at finite chemical potential ($ \mu >0$). 

The pole exhibits a dispersion relation identical to that of the Landau's ``zero sound'' whose real part is linear in momentum $ k $, while the imaginary part goes quadratic ($\sim k^2 $) in momentum.  In a weakly coupled system, such ``collective excitations'' are naturally identified with Landau's zero sound that are associated with fluctuations near the Fermi surface \cite{L1}-\cite{b2}. However, such an interpretation becomes obscure in a holographic set up that deals with SCFTs with much richer structure, containing both fermions and bosons. In theories with Lifshitz scaling ($ z>1 $), the Green's function exhibits a quasiparticle peak for $ z<2 $, that can be identified with zero sound modes \cite{Hoyos-Badajoz:2010ckd}, following an analogy with Landau's theory on Fermi liquids. The interesting fact about holographic calculations is the presence of the zero sound modes despite of the non-Fermi or non-Bose liquid type behaviour of the heat capacity at low temperatures \cite{Karch:2008fa}.

The purpose of the present paper is to investigate the above possibilities for $ \mathcal{N}=2 $ SCFTs in 4d \cite{Gaiotto:2009we} that are dual to Gaiotto-Maldacena (GM) class of geometries in 10d \cite{Gaiotto:2009gz}. In particular, in our analysis, we focus on a particular class of solution in the GM category namely, the non-Abelian T-dual (NATD) of $ AdS_5 \times S^5 $  \cite{Lozano:2016kum}. The two point correlations in the color and the flavour sectors are studied separately by adding D-branes as a ``probe'' over NATD background. 

For our purpose, we introduce $ N_4(\ll N_c) $ probe D4 branes that source ``color carriers'' and $ N_f (\ll N_c)$ probe D6 branes that source ``flavour carriers'' for the dual $ \mathcal{N}=2 $ quiver \cite{Lozano:2016kum}, where $ N_c( \rightarrow \infty )$ corresponds to number of color nodes in the super-gravity limit. When D4 world-volume $ U(1) $ is turned on, it sources a $ U(1) $ charge density ($ J^t_c \sim \varrho_c $) for the ``color carriers''  of $ \mathcal{N}=2 $ quiver. In other words, the color carriers of $ \mathcal{N}=2 $ quiver are charged under the world-volume $ U(1) $ of D4, that results in a non-vanishing chemical potential ($ \mu_c $). 

On the other hand, when $ N_f(\ll N_c )$ D6 are added, they correspond to $ N_f $ flavours that are in the fundamental of $ SU(N_c) $ gauge group of $ \mathcal{N}=2 $ quiver. In this paper, we would be interested in the ``quasiparticle'' excitations that are charged under the world-volume $ U(1) $ of D6. The D6 world-volume gauge field triggers a ``flavour'' charge density ($ J_f^t \sim \varrho_f $) and hence an associated flavour chemical potential ($ \mu_f $) for the dual $ \mathcal{N}=2 $ SCFTs at $ T=0 $. 

The set up above, provides a perfect platform to calculate correlation functions at finite density and explore its pole at low frequency ($ \omega $) and low momentum ($ k $). We have two types of probe D-branes and depending on their nature, one can study correlators either in the color or in the flavour sector of $ \mathcal{N}=2 $ quiver. For the purpose of this paper, we are interested in the charge-charge and current-current retarded correlators and their low frequency behaviour.

While performing these calculations for the color sector, our analysis reveals the existence of ``gapless'' excitations that can be broadly categorised into two classes. One of these quasiparticle excitations move with a constant phase velocity, while the other class of collective excitations quantum attenuates due to interactions with the ``color'' bath. These second class of quantum modes exhibit a dispersion relation in which the imaginary part of the frequency goes linearly with the momentum, which therefore rules out the possibilities of holographic zero sound for the color sector of $ \mathcal{N}=2 $ quiver. We also compute the AC conductivity ($ \sigma(\omega) $) at low frequencies, which turns out to be purely imaginary that exhibits a simple pole at vanishing frequency.

Next, we repeat above calculations for $N_f(\ll N_c)$ flavour D6 branes. We choose to work with a particular embedding namely, one of the world-volume directions of D6 is extended along the holographic $r$-axis while the two among the remaining directions wrap the internal $S^2$ of the Sfetsos-Thompson (ST) background \cite{Lozano:2016kum}. It turns out that the associated (low energy) dispersion relation could be categorised into two classes -(i) the one that contains quasiparticle excitation which moves with a constant phase velocity and with no quantum attenuation and (ii) the other that propagates with a definite ``group velocity" and quantum attenuates similar in spirit as that of the Landau's zero sound mode. We also calculate the AC conductivity associated with the flavour sector, which exhibits a higher order pole at low frequencies. 

The rest of the paper is organised as follows. In Section 2, we explore $ \mathcal{N}=2 $ quivers at finite density and estimate the speed of the first sound both in the color and the flavour sectors. In Section 3, we study low frequency behaviour of holographic two point function(s) in the color sector. We work out the dispersion relation by looking into the pole(s) appearing in the two point correlator(s). We perform similar analysis for the flavour sector in Section 4. Finally, we draw our conclusion in Section 5, along with some future remarks and possible extensions.
\section{ $\mathcal{N}=2$ quivers at finite density}
We begin by introducing type IIA backgrounds that are obtained via non-Abelian T- duality (NATD) along $ SU(2)\subset S^5 $. The metric in the string frame reads as \cite{Lozano:2016kum}
\begin{align}
\label{e2.1}
&ds_{10}^2 = \frac{4R^2}{L^2}dx^2_{1,3}+\frac{4L^2}{R^2}dR^2+4L^2(d \alpha^2 + \sin^2\alpha d\beta^2)+\frac{\alpha'^2}{L^2 \cos^2\alpha}dr^2\nonumber\\
&+\frac{\alpha'^2 L^2 \cos^2\alpha r^2}{\alpha'^2 r^2+L^4 \cos^4\alpha}(d\chi^2 + \sin^2\chi d\xi^2)\\
&B_2 = \frac{\alpha'^3 r^3}{\alpha'^2 r^2+L^4 \cos^4\alpha}\sin\chi d\chi \wedge d\xi ~;~e^{-2\Phi}=\frac{L^2}{\alpha'^3}\cos^2\alpha ({\alpha'^2 r^2+L^4 \cos^4\alpha})\\
&F_2=dC_1=\frac{8L^4}{\alpha'^{3/2}}\sin\alpha \cos^3\alpha d\alpha \wedge d\beta ~;~F_4=dC_3=B_2 \wedge F_2.
\label{e2.3}
\end{align}

The above background\footnote{Also known as the Sfetsos-Thompson (ST) solution  \cite{Sfetsos:2010uq}.} \eqref{e2.1}-\eqref{e2.3} preserves $ \mathcal{N}=2 $ SUSY \cite{Sfetsos:2010uq}. Notice that, the metric and the dilaton are singular at $ \alpha = \frac{\pi}{2} $. Following \cite{Lozano:2016kum}, we take the $ r $ coordinate to be varying in the interval $ [0,n \pi] $, where $ n $ is a positive integer. In our analysis, $ n $ is considered to be large number that stands for the ``rank'' of the color $SU(N_c)$ gauge group of $\mathcal{N}=2$ quiver. In what follows, we estimate correlators in a large rank ($ n \rightarrow \infty $) limit of the $ SU(N_c) $ color group, that takes care of the cumulative effects due to all color nodes pertaining to $ \mathcal{N}=2 $ quiver.

\begin{figure}
\begin{center}
\includegraphics[scale=0.45]{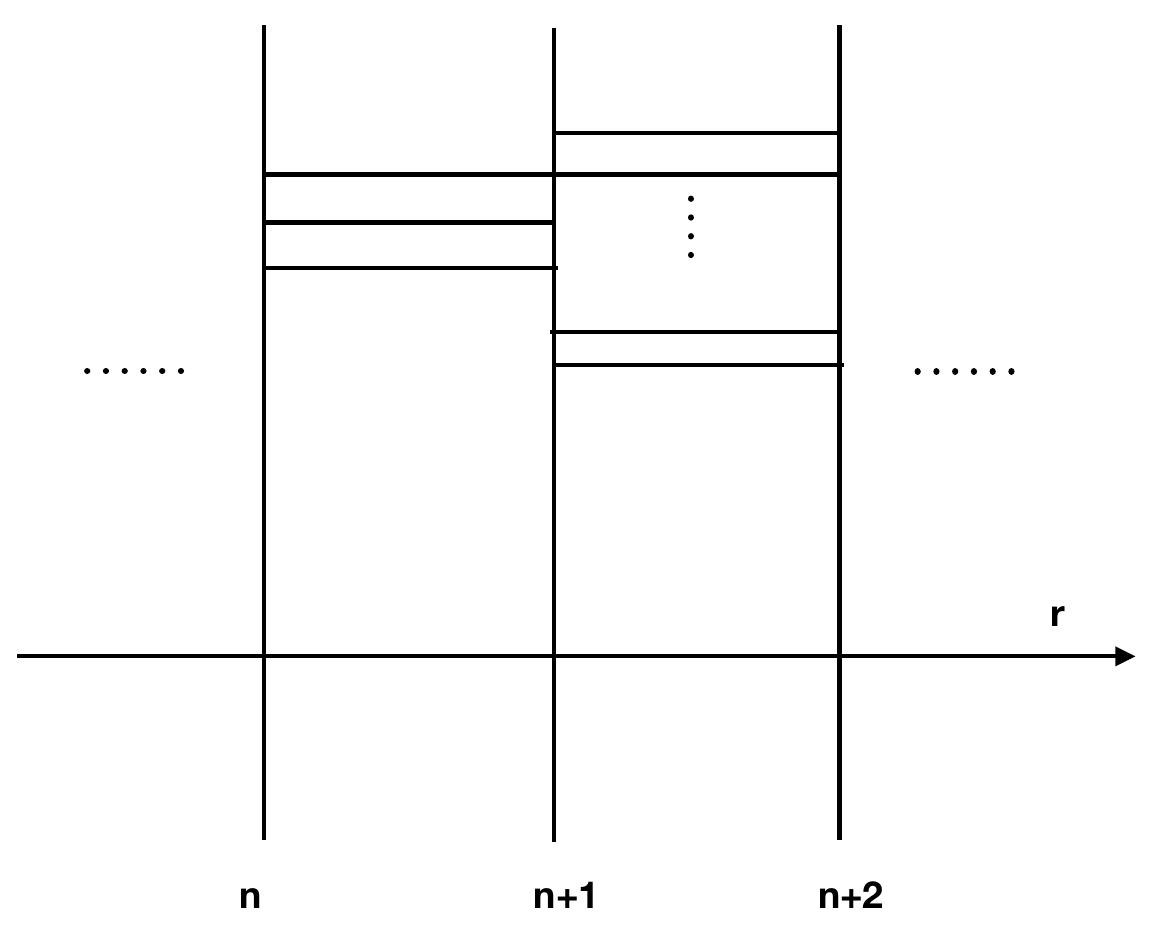}
  \caption{Hanany-Witten setup that realises $ \mathcal{N}=2 $ SCFTs in 4d. In the figure above, NS5 branes are separated in the units of $ \pi $. For each of these intervals, we have color D4 stretched between two NS5. In the supergravity/ strong coupling limt of the quiver, one of these world-volume directions goes to zero, which effectively gives rise to $ \mathcal{N}=2 $ SCFTs living in 4d.} \label{fig}
  \end{center}
\end{figure}

It turns out that the background \eqref{e2.1}-\eqref{e2.3} falls under the general category of GM class of solutions \cite{Lozano:2016kum}. The corresponding Hanany-Witten set up comprises of NS5 branes that are equally spaced along the holographic $r$- axis, while $N_c$ color D4 branes are stretched between them. World-volume theories of these color branes are characterised by $SU(N_c)$ gauge group. 

The number of color D4 in each interval $[n \pi, (n+1)\pi] $ increases by one unit in $ N_c $, which thereby increases the rank of the associated color group by unity. The above procedure results in a $ \mathcal{N}=2 $ linear quiver with linearly increasing rank. The rank eventually increases upto infinity starting from zero, which implies that the $ \mathcal{N}=2 $ quiver is of infinite length. The super-gravity description is a good approximation in the limit when all NS5 branes are put on top of each other together with the fact that $ N_c \rightarrow \infty $. This is the limit, in which the coupling of $ \mathcal{N}=2 $ quiver grows to infinity and the world-volume theory is ``effectively'' four dimensional \cite{Lozano:2016kum} as is also mentioned earlier. The analysis of the present paper always confines to this limiting scenario, where the dual superconformal quiver is strongly coupled which has low energy fluctuations\footnote{By low energy excitations, we refer to the fact that the holographic two point functions in Section \ref{color} and Section \ref{flavour} are estimated in the limit of zero frequency ($ \tilde{\omega}\rightarrow 0 $) and zero momentum ($ \tilde{k}\rightarrow 0 $). These correlators are due to external charge carriers that are added through probe D- branes and behave much like that of quasi-particle excitation over a strongly coupled ground state in the dual superconformal quiver. Bringing an analogy with standard condensed matter systems, these charge carriers are similar to those of the low energy quasiparticle excitations near the Fermi surface, which is the groundstate of a system of fermions at $T=0$.} that behave like quasiparticle excitations. In the dual super-gravity description \eqref{e2.1}, it is the holographic $ r $- axis that takes care of this linearly increasing rank function of the quiver and in principle ranges between zero to infinity. 
\subsection{Probe D4 branes}
We consider $ N_4 (\ll N_c)$ probe D4 branes \cite{Lozano:2016kum} that are extended along $ \mathbb{R}^{1,3}\times r $ and are placed at a fixed radial coordinate $ R $ of the $ AdS_5 $. The corresponding world-volume action is given by
\begin{align}
\label{e2.4}
&-S_{D4}=N_4 T_{D4}\int d^5 x \Big[  e^{-\Phi}\sqrt{-\det (g_{MN}+2\pi \alpha' \mathcal{F}_{MN})} -C_5 \Big]\\
&C_5 = 16\sqrt{\alpha'}\frac{R^4}{L^4}r dt \wedge dx_1 \wedge dx_2 \wedge dx_3 \wedge dr.
\end{align}
Here, $ \mathcal{F}_{MN} =\partial_M \mathcal{A}_N - \partial_N \mathcal{A}_M$ is the world-volume field strength tensor. For our purpose, we choose $ \mathcal{A}_M=(\mathcal{A}_t(r),0,0,0,0) $. This implies that the only non-vanishing component of the field strength tensor is $ \mathcal{F}_{t r}=-\mathcal{A}_t' $. Using this, one could simplify \eqref{e2.4} further to yield
\begin{align}
-S_{D4}=N_4 T_{D4}\frac{16\sqrt{\alpha'}R^4}{L^4}\text{Vol}(\mathbb{R}^{1,3})\int_0^{n \pi}dr  \Big[f(r)\sqrt{1-\frac{\pi^2 L^4}{R^2}\mathcal{A}_t'^2} -r\Big]
\end{align}
where we define $ f(r)=\sqrt{r^2+\frac{L^4}{\alpha'^2}} $ and set $ \alpha=0 $ as an initial choice that preserves SUSY. 

The equation of motion for the $ U(1) $ gauge field reads 
\begin{align}
\label{e2.7}
\mathcal{A}_t'=\frac{C}{f(r)}\frac{1}{{\sqrt{1+\frac{\pi^2 L^4 C^2}{R^2 f^2}}}}
\end{align}
where $ C $ is a dimensionful ($ [L]^{-1} $) constant of integration. The above equation \eqref{e2.7} can be integrated to solve for $ \mathcal{A}_t $. World-volume $ U(1) $ gauge field of the probe D4 brane sources a (color) chemical potential ($ \mu_c $) and a (color) charge density ($ \varrho_c $) for the dual $ \mathcal{N}=2 $ quiver \cite{Hartnoll:2009ns} at strong coupling. The chemical potential can be obtained using the following expression\footnote{Notice that, here $ r $ is the direction that takes care of all the color nodes and is not the radial direction in the bulk $ AdS_5 $. The probe D4-brane is placed at a fixed radial distance $ R $ and it expands along all other spatial axes including the quiver direction $ r $. One should think of the distance $ R $ as the location of the boundary where the dual superconformal quiver lives. Given the above picture, one can now integrate over the direction $ r $ in order to find a cumulative effect on boundary observables due to all color nodes pertaining to the quiver. As we integrate over all possible color nodes (which in principle sets $ n $ to be very large), the contribution to the chemical potential due to each node adds up and results in a large number which is depicted in \eqref{e2.9}.} 
\begin{align}
\label{e2.8}
\mu_c = \int_0^{n \pi}dr \mathcal{A}'_t =\frac{C}{2}\log\Big[\frac{4 \pi^2 \alpha'^2 n^2}{L^4 (1+\frac{\pi^2 C^2 \alpha'^2}{R^2})} \Big]+\frac{C L^4 }{4 n^2}\left(\frac{1}{\pi ^2 \alpha'^2}+\frac{C^2}{R^2}\right)+\mathcal{O}(1/n^3).
\end{align}

One can simplify \eqref{e2.8} by setting constants in an appropriate manner $ 1+\frac{\pi^2 C^2 \alpha'^2}{R^2}=\frac{4\pi^2 \alpha'^2}{L^4} $, which reveals a simplified expression of the form
\begin{align}
\label{e2.9}
\mu_c = C \log n+\mathcal{O}(1/n^2).
\end{align}
Clearly, in the limit of large $ n$, the color chemical potential goes like $ \mu_c \sim \log n \sim \log N_5 $ where $ N_5 $ is the number of NS5 branes associated with the NATD background in the interval $ [0, n \pi] $. This is the ``Abelian'' limit of the NATD where the coupling of the D4 brane becomes small \cite{Lozano:2016kum}.

Using \eqref{e2.7}, the grand canonical free energy can be obtained from on-shell action \eqref{e2.4}
\begin{align}
\label{e2.10}
\Omega_c = -S_{D4}=N_4 T_{D4}\frac{16\sqrt{\alpha'}R^4}{L^4}\text{Vol}(\mathbb{R}^{1,3})\int_0^{n \pi}dr \Big[ \frac{f^2(r)}{\sqrt{f^2(r)+\frac{\pi^2 L^4 C^2}{R^2}}}-r \Big].
\end{align}

The integral \eqref{e2.10} can be obtained as an expansion in the large $ n $ limit
\begin{align}
\label{e2.11}
\Omega_c = \Omega_0+\frac{(\Omega_0-\pi^2)}{\log 2}\log(\frac{n}{2})+\mathcal{O}(1/n^2)
\end{align}
where $ \Omega_0=\pi^2 (1+\frac{L^4}{4 \pi^2 \alpha'^2}\log4(1-\frac{\pi^2 \alpha'^2 C^2}{R^2})) $ is a constant and we re-scale $ \Omega_c $ by the overall pre-factor in \eqref{e2.10} that includes the Minkowski volume $ \text{Vol}(\mathbb{R}^{1,3}) $.

Given the free energy \eqref{e2.11}, charge density follows from thermodynamic identities
\begin{align}
\varrho_c = -\frac{\partial \Omega_c}{\partial \mu_c}=-\frac{(\Omega_0-\pi^2)}{C\log 2}+\mathcal{O}(1/n^2).
\end{align}

The energy density ($ \varepsilon_c $) and pressure $ p_c (=-\Omega_c)$ can be related following standard thermodynamic identities in a grand canonical ensemble
\begin{align}
\label{e2.13}
\varepsilon_c =  \Big(  \frac{\mu_c}{C \log (\frac{n}{2})}-1\Big)p_c+\frac{\mu_c}{C}\frac{\Omega_0}{\log (\frac{n}{2})}+\mathcal{O}(1/n^2).
\end{align}

Using \eqref{e2.13}, the speed of sound \cite{Hoyos-Badajoz:2010ckd}-\cite{Pang:2013ypa} in the grand canonical ensemble turns out to be
\begin{align}
\label{e2.14}
\upsilon_c^2 = \Big|\frac{\partial p_c}{\partial \varepsilon_c}\Big|_{\mu_c}=\Big| 1- \frac{\mu_c}{C \log (\frac{n}{2})}\Big|^{-1}.
\end{align}

Clearly, one finds that for a given chemical potential ($ \mu_c $), the sub-leading corrections appearing in \eqref{e2.14} are suppressed by the inverse power of the logarithm of the rank function of the quiver. Furthermore, a careful look reveals that in the Abelian ($ n\rightarrow \infty $) limit, the numerator and the denominator both are of the order $\sim \log n $, which therefore produces a non zero contribution only at sub-leading order in the $ 1/n $ expansion. 

Notice that, in the above analysis, we set $\alpha =0$ for simplicity. This is an allowed choice since the metric \eqref{e2.1} is well defined. On the other hand, with the choice $ \alpha =\frac{\pi}{2} $, the metric \eqref{e2.1} blows up and therefore should be discarded.
\subsection{Probe D6 branes}
When $ N_f (\ll N_c) $ D6 branes are added, it source $ N_f $ flavour multiplet for the $ \mathcal{N}=2 $ quiver. These are what we denote as flavour degrees of freedom which are ``massless'' for the present model that we consider below\footnote{Massive charge carriers are introduced by incorporating an appropriate embedding (scalar) field on the world-volume of the probe D- brane that acts as dual to the flavour mass operator. In other words, one introduces an embedding function for the probe D-brane, which acts as a cyclic variable for the DBI Lagragian \cite{Karch:2007br}-\cite{Lee:2010uy}.}. The corresponding DBI action reads as
\begin{align}
-S_{D6}/N_f= T_{D6}\int d^7x  e^{-\Phi}\sqrt{-\det (g_{MN}+B_{MN}+2\pi \alpha' \mathcal{F}_{MN})} -T_{D6}\int C_7 -C_5 \wedge B_2
\end{align}
where we choose the world-volume $ U(1) $ field strength as before, $ \mathcal{F}_2=-\mathcal{A}_t'(r)dt \wedge dr$. Turning on $ U(1) $ on D6 triggers a ``flavour current'' ($ J_f^\mu $) and hence a flavour charge density ($ \varrho_f $) for the dual $ \mathcal{N}=2 $ quiver. In other words, these ``flavour carriers'' are charged under $ U(1) $ that corresponds to adding a ``flavour'' chemical potential ($ \mu_f $) for the dual $ \mathcal{N}=2 $ quiver.

We consider that the D6 brane world-volume directions are extended along $ \mathbb{R}^{1,3}\times r\times S^2(\chi ,\xi) $ and it is placed at a fixed radial distance $ R $ of the $ AdS_5 $. A straightforward calculation reveals the following world-volume action for the $ N_f $ D6 brane
\begin{align}
\label{e2.17}
-S_{D6}/N_f=T_{D6}\frac{64 \pi \alpha'^{3/2} R^4}{L^4}\text{Vol}(\mathbb{R}^{1,3})\int_0^{n \pi}dr  \Big[r^2\sqrt{1-\frac{\pi^2 L^4}{R^2}\mathcal{A}_t'^2} -r^2\Big]
\end{align}
where we set $ \alpha =0 $ without any loss of generality. 

The equation of motion for the $ U(1) $ gauge field reads
\begin{align}
\label{e2.18}
\mathcal{A}_t'=\frac{C}{r^2}\frac{1}{{\sqrt{1+\frac{\pi^2 L^4 C^2}{R^2 r^4}}}}.
\end{align}

The chemical potential in the flavour sector can be obtained following the integration
\begin{align}
\mu_f = \int_0^{n \pi}dr \mathcal{A}'_t = \frac{n R}{L^2}\, _2F_1\left(\frac{1}{4},\frac{1}{2};\frac{5}{4};-\frac{\pi ^2 n^4 R^2}{C^2 L^4}\right).
\end{align}

By expanding in the large $ n $ limit, one finds
\begin{align}
\label{e2.20}
\mu_f =\bar{ \mu}-\frac{C}{\pi n}+\mathcal{O}(1/n^5)
\end{align}
where we identify, $\bar{ \mu}=\frac{\sqrt{C R}}{4\pi L} \Gamma^2(\frac{1}{4})$. The next step is to obtain the free energy ($ \Omega_f $) associated with the flavour modes in a grand canonical ensemble. This follows by substituting the solution \eqref{e2.18} into \eqref{e2.17} and performing a proper rescaling by $ T_{D6}\frac{64 \pi \alpha'^{3/2} R^4}{L^4}\text{Vol}(\mathbb{R}^{1,3}) $
\begin{align}
\label{e2.21}
\Omega_f = -\frac{1}{3} \pi ^2 n \left(\sqrt{\frac{C^2 L^4}{R^2}+\pi ^2 n^4} \, _2F_1\left(\frac{3}{4},1;\frac{5}{4};-\frac{n^4 \pi ^2 R^2}{C^2 L^4}\right)-\sqrt{\frac{C^2 L^4}{R^2}+\pi ^2 n^4}+\pi  n^2\right).
\end{align}

Considering a large $ n $ limit, the above expression \eqref{e2.21} simplifies further to yield
\begin{align}
\label{e2.22}
\Omega_f =-\Omega_0+\frac{\pi C^2 L^4}{2R^2 n}+\mathcal{O}(1/n^5)
\end{align}
where $ \Omega_0= \frac{\pi}{12}\Gamma^2(\frac{1}{4})(\frac{C^2 L^4}{R^2})^{3/4}$. Using \eqref{e2.20} and \eqref{e2.22}, one can further estimate the charge density ($ \varrho_f $) in the flavour sector within the grand canonical framework
\begin{align}
\varrho_f =-\frac{\partial \Omega_f}{\partial \mu_f}=\frac{\pi^2 L^4 C}{2R^2}+\mathcal{O}(1/n^4).
\end{align}

The energy density ($ \varepsilon_f $) in the flavour sector can be expressed as
\begin{align}
\varepsilon_f = \frac{\pi \bar{\mu} n}{C}\Omega_f+\frac{\mu_f \pi n}{C}\Omega_0
\end{align}
which can be used to estimate the sound speed ($ \upsilon_f $) in a grand canonical ensemble
\begin{align}
\label{e2.24}
\upsilon^{2}_f =\frac{C}{\pi \bar{\mu}n}=\frac{4L\sqrt{C}}{\sqrt{R}\Gamma^2(\frac{1}{4})}\frac{1}{n}.
\end{align}
 
One could see that the speed of sound \eqref{e2.24} decreases with increasing rank ($n$) of the color gauge group in the dual quiver. This stems from the fact that flavour carriers interact with the ``color bath''  and are scattered. This causes an overall damping of sound modes in the flavour sector. With increasing rank, the scattering and hence the effects of damping is enhanced. 
\section{Correlations at finite density : color sector}
\label{color}
We are interested in computing two types of correlators for $ \mathcal{N}=2 $ quiver at $ T=0 $. These are the retarded density-density ($ G^{tt}_R (\omega , k) $) and current-current ($ G^{xx}_R (\omega , k) $) two point functions. As shown by the authors in \cite{Karch:2008fa}-\cite{Pang:2013ypa}, the pole appearing in the density two point function ($ G^{tt}_R (\omega , k) $) at low frequency ($ \omega $) and momentum ($ k $) gives the zero sound. On the other hand, a similar calculation for the current-current two point function ($ G^{xx}_R (\omega , k) $) yields the AC conductivity following Kubo's formula. In this section, we perform a similar computation for the color sector and then we repeat these calculations for the flavour modes in the next section. 

While computing two point functions in the color sector, we do not anticipate anything a priori about the low frequency behaviour of the retarded correlators. The on-shell action acts like a generating functional for the dual $ \mathcal{N}=2 $ SCFT. We turn on $ \mathcal{A}_t $ and $ \mathcal{A}_x $ gauge field components on the D4 brane world-volume that source density ($ J^t_c $) and current operator ($ J^x_c $) for the dual super-conformal theory at strong coupling. We solve corresponding world-volume fluctuations at linearised order, which when substituted back yields the desired on-shell action.

We consider fluctuations of the form $ \mathcal{A}_\mu \rightarrow \mathcal{A}_\mu (r)+a_\mu (t, r, x) $, where we exploit translational invariance to choose one of the spatial directions as $ x_1 =x$. Expanding \eqref{e2.4} upto quadratic order in the fluctuations ($ a_\mu  $), one finds the quadratic world-volume action
\begin{align}
\label{e3.1}
&-S^{(2)}_{D4}=N_4 T_{D4}\text{Vol}(\mathbb{R}^{1,3})\int_0^{n \pi}dr \frac{L}{\sqrt{\alpha'}}f(r)\Big( \frac{8\pi^2 R^2 \alpha' }{L}\frac{f^2_{xr}}{\sqrt{1-\frac{\pi^2 L^4}{R^2}\mathcal{A}_t'^2}}\nonumber\\
& -\frac{2 \pi^2 \alpha'^3}{L}\frac{f^2_{tx}}{\sqrt{1-\frac{\pi^2 L^4}{R^2}\mathcal{A}_t'^2}}-\frac{8 \pi^2 R^2 \alpha'}{L}\frac{f^2_{tr}}{\Big( 1-\frac{\pi^2 L^4}{R^2}\mathcal{A}_t'^2\Big)^{3/2}} \Big)
\end{align}
where we denote $ f_{MN}=\partial_{[M} a_{N ]}$.

We solve these fluctuations in the Fourier space, where we introduce Fourier transformation
\begin{align}
\label{e3.2}
a_\mu (t, r, x)=\int \frac{d \omega dk}{(2\pi)^2}e^{-i\omega t +i k x}a_\mu (r, \omega , k)
\end{align}
where we choose to work with the gauge, $ a_r = 0 $ \cite{Karch:2008fa}-\cite{Pang:2013ypa}.

The equations of motion at the linearised level take the following form
\begin{align}
\label{e3.3}
&\partial_r (g(r)a_x')-\frac{\alpha'^2}{4 R^2}g(r)(\omega^2 a_x + \omega k a_t)=0\\
\label{e3.4}
&\partial_r \Big(\frac{g^3(r)}{f^2(r)}a_t'\Big)+\frac{\alpha'^2}{4 R^2}g(r)(\omega k a_x +  k^2 a_t)=0\\
&k a_x' +\frac{g^2}{f^2}\omega a_t' =0
\label{e3.5}
\end{align}
where prime stands for the derivative with respect to $ r $ and $ g(r)= \sqrt{f^2(r)+\frac{\pi^2 L^4 C^2}{R^2}}$. The last equation \eqref{e3.5} is an artefact of the $ a_r $ equation of motion expressed in $ a_r=0 $ gauge. One can substitute $ a_x' $ from \eqref{e3.5} into \eqref{e3.3} to show that it precisely reproduces \eqref{e3.4}, which therefore can be ignored at the level of the dynamics. To summarise, we solve only equations \eqref{e3.4}-\eqref{e3.5}.

Next, we introduce a gauge invariant observable $ E(r, \omega, k)=\omega a_x +k a_t $, which when substituted back into \eqref{e3.5} yields the following set of solutions
\begin{align}
\label{e3.6}
a_t' = \frac{f^2(r)k}{f^2(r)k^2 - g^2(r)\omega^2}E' ~;~ a_x' =-\frac{g^2(r)\omega}{f^2(r)k^2 - g^2(r)\omega^2}E'.
\end{align}

Substituting \eqref{e3.6} into \eqref{e3.4}, we find the equation of motion for $ E(r, \omega , k) $
\begin{align}
\label{e3.7}
E'' + \partial_r \log \Big( \frac{g^3(r)}{f^2(r)k^2-g^2(r)\omega^2} \Big)E'+\frac{\alpha'^2}{4R^2}\Big( \frac{f^2 (r)}{g^2(r)}k^2-\omega^2 \Big)E=0.
\end{align}

The corresponding quadratic action \eqref{e3.1} turns out to be
\begin{align}
-S^{(2)}_{D4}=\mathcal{N}_{D4}\int dr d\omega dk g(r)\Big( E^2-\frac{4R^2}{\alpha'^2}\frac{g^2(r)E'^2 }{(f^2(r)k^2 - g^2(r)\omega^2)}\Big)
\end{align}
where we introduce the pre-factor as, $ \mathcal{N}_{D4}= 2\pi^2 N_4 T_{D4}\alpha'^{5/2}\text{Vol}(\mathbb{R}^{1,3})$.

Using \eqref{e3.7}, the ``on-shell'' action can be further simplified as
\begin{align}
\label{e3.9}
S^{(2)}_{D4}\Big|_{\partial r}:=\mathcal{Z}_{c}=\frac{4R^2}{\alpha'^2}\mathcal{N}_{D4}\int d\omega dk \frac{g^3 (r)E'(r)E(r)}{(f^2(r)k^2 -g^2(r) \omega^2)}\Big|^{r= n \pi}_{r= 0}
\end{align}
which we identify as the partition function for the color nodes at finite $ U(1) $ density of color carriers in $ \mathcal{N}=2 $ SCFTs. One can further simplify \eqref{e3.9} considering a large $ n $ limit
\begin{align}
\label{e3.10}
\mathcal{Z}_{c}=\frac{4R^2}{\alpha'^2}\mathcal{N}_{D4}\Big[n \pi \int  \frac{d\omega dk}{(k^2 - \omega^2)}E'(r)E(r)\Big|_{r=n\pi}-\int  \frac{\zeta^{3/2}d\omega dk}{(\frac{L^4}{\alpha'^2} k^2- \zeta \omega^2)}E'(r)E(r)\Big|_{r=0}\Big]
\end{align}
where $ \zeta =\frac{L^4}{\alpha'^2} + \frac{\pi ^2 L^4 C^2 }{R^2} $ is a dimensionless constant. A closer look further reveals that in the limit of low frequency and momenta, the solution of \eqref{e3.7} takes the form, $ E(r)\sim r+ \mathcal{O}(k^2,\omega^2) $. In other words, the second integral in \eqref{e3.10} yields a vanishing contribution to $ \mathcal{Z}_{c} $.

We estimate two point charge-charge and current-current correlators of the form
\begin{align}
\label{e3.11}
\mathcal{G}^{(R)}_{tt}(\omega , k)=k^2 \Pi (\omega , k)~;~\mathcal{G}^{(R)}_{xx}(\omega , k)=\omega^2 \Pi (\omega , k)
\end{align}
in the regime of low frequency ($  \frac{\omega \alpha'}{R} \ll 1 $) and low momenta ($ \frac{k \alpha'}{R} \ll 1 $), where we identify
\begin{align}
\label{e3.12}
\Pi (\omega , k):=\frac{\delta^2}{\delta E (r)^2}\mathcal{Z}_{c}\Big|_{r= n \pi}.
\end{align}

Following \cite{Karch:2008fa}-\cite{Pang:2013ypa}, we solve \eqref{e3.7} in two steps. In the first approach, we take a large $ r(=n \pi) $ limit first and thereby expanding the solution $ E(r) $ in the regime of small frequency and momenta as mentioned above. This yields a differential equation of the following form
\begin{align}
E''(r)+\frac{1}{r}E'(r)+\frac{\alpha'^2}{4R^2}(k^2 - \omega^2)E(r) =0 
\end{align}
which has a solution in the form of Bessel's functions of first kind
\begin{align}
\label{e3.14}
E(r,\tilde{\omega}, \tilde{k})=c_1 J_0\left(\frac{r  \sqrt{\tilde{k}^2-\tilde{\omega} ^2}}{2}\right)+c_2 Y_0\left(\frac{r \sqrt{\tilde{k}^2-\tilde{\omega} ^2}}{2}\right)
\end{align}
where we introduce new dimensionless entities, $ \tilde{k}=\frac{k \alpha'}{R} $ and $ \tilde{\omega}=\frac{\omega \alpha'}{R} $.

Out of these two solutions, the solution with coefficient $ c_2 $ yields an incoming wave \cite{Son:2002sd}-\cite{vanRees:2009rw} $ E(r)\sim \frac{1}{\sqrt{r}} \exp \left(\pm\frac{1}{2} i \sqrt{\tilde{k}^2-\tilde{\omega}^2} r\right)$ as $ r\rightarrow \infty $, which is therefore the solution of our interest \cite{Karch:2008fa}. Notice that, unlike the previous studies \cite{Karch:2008fa}-\cite{Pang:2013ypa}, the function \eqref{e3.14} depends both on the frequency ($ \tilde{\omega} $) and momentum ($ \tilde{k} $). In principle, therefore there are two possible ways to expand the function \eqref{e3.14}. However, a closer look further reveals that this problem can be evaded if we first consider an expansion using the ratio $ \frac{\tilde{\omega}}{\tilde{k}}$ and then an expansion in the regime of small momentum ($ \tilde{k} r \ll 1$). This leads to the following expression\footnote{Here, $ \gamma_E $ is the Euler's gamma function.}
\begin{align}
\label{e3.15}
E (\tilde{\omega} ,\tilde{k})\Big|_{r=n \pi} = -\frac{\tilde{C} \tilde{\omega}^2}{\pi \tilde{k}^2}+\frac{2\tilde{C}}{\pi} \left(\log (n \pi)-2 \log 2+\log \tilde{k}+\gamma_E \right)
\end{align}
which is subjected to the constraint, $ \tilde{k}^2 - \tilde{\omega}^2=\frac{16}{\pi^2 n^2} \ll 1$. Notice that, $ \log \tilde{k} $ diverges in the limit of small momentum. When substituted back into \eqref{e3.10}, this yields a divergent contribution $ \sim \int \frac{d \tilde{k}}{\tilde{k}^2}\log \tilde{k} $ that can be absorbed by adding a suitable counter term in \eqref{e3.10}. In other words, one can remove $ \log \tilde{k} $ in \eqref{e3.15} while considering a ``regularised'' partition function. 

In the second approach, we first expand \eqref{e3.7} in small frequency ($ \tilde{\omega} \ll 1$) and then in small momentum ($ \tilde{k}\ll1 $), such that the ratio $ \frac{\tilde{\omega}}{\tilde{k}}\ll 1 $. This leads to the following equation
\begin{align}
E''(r)+\Big(\frac{1}{r}+\frac{L^4 \left(\omega ^2 \left(\pi ^2 \alpha'^2 C^2+R^2\right)-k^2 \left(3 \pi ^2 \alpha'^2 C^2+R^2\right)\right)}{\alpha'^2 r^3 R^2 \left(k^2-\omega ^2\right)}\Big)E'(r)=0.
\end{align}

The corresponding solution can be expressed using the exponential integral function $  \text{Ei}(x) $
\begin{align}
\label{e3.17}
E(r, \tilde{\omega}, \tilde{k})=C_2-\frac{C_1}{2}  \text{Ei}\left(-\frac{L^4 \pi^2 C^2R^2 \left(\tilde{k}^2 \left(3+\frac{R^2}{\pi^2 \alpha'^2 C^2}\right)-\left(1+\frac{R^2}{\pi^2 \alpha'^2 C^2}\right) \tilde{\omega}^2\right)}{2 r^2 R^4 \left(\tilde{k}^2-\tilde{\omega} ^2\right)}\right).
\end{align}

Setting $ r= n \pi $ and thereby taking a large $ n $ limit, we find
\begin{align}
\label{e3.18}
E (\tilde{\omega} ,\tilde{k})\Big|_{r=n \pi} =C_2-\frac{\gamma_E  C_1}{2}+C_1 \log (n \pi)+\frac{C_1}{2}\log\Lambda-\frac{\pi^2 \tilde{\omega}^2L^4 C^2 C_1 \Lambda}{2\tilde{k}^2R^2}+\mathcal{O}(\tilde{\omega}^4/\tilde{k}^4)
\end{align}
which correctly reproduces the $ \log (n \pi) $ behaviour that is expected from \eqref{e3.15}.

Following \cite{Hoyos:2011zz}, we can fix the constants $ C_1 $ and $C_2 $ in terms of $ \tilde{C} $ by equating the coefficients of constant terms and $ \log (n \pi) $ between \eqref{e3.15} and \eqref{e3.18}. This leads to the following solutions, $ C_1 = \frac{2\tilde{C}}{\pi} $ and $ C_2= \frac{2\tilde{C}}{\pi}(\frac{3}{2}\gamma_E - \log(4 \sqrt{\Lambda}))$, where $ \Lambda=  \frac{2 \alpha'^2}{L^4(1+\frac{3 \pi^2 \alpha'^2 C^2}{R^2})} $ is a dimensionless entity. Finally, by comparing the coefficients of $ \frac{\tilde{\omega}^2}{\tilde{k}^2} $ in \eqref{e3.15} and \eqref{e3.18}, we find $ \frac{L^4 C^2 \Lambda}{R^2}=\frac{1}{\pi^2} $.

Using the above solutions and \eqref{e3.17}, the partition function \eqref{e3.10} becomes
\begin{align}
\label{e3.19}
\mathcal{Z}_{c}=\frac{R^2}{\alpha'^2}\mathcal{N}_{D4}\int  \frac{d\tilde{\omega} d\tilde{k}}{(\tilde{k}^2 - \tilde{\omega}^2)}\Sigma (\tilde{\omega}, \tilde{k})+\mathcal{O}(1/n^2)
\end{align}
where the function within the integral can be identified as
\begin{align}
\label{e3.20}
\Sigma (\tilde{\omega}, \tilde{k})=\tilde{C}^2 \Big( \frac{16}{\pi^2}\gamma_E -\frac{8 \tilde{\omega}^2 }{\pi^2\tilde{k}^2 } +\frac{16}{\pi^2}\log(\frac{n \pi}{4})\Big).
\end{align}

One should notice that the above expression \eqref{e3.20} and hence the partition function \eqref{e3.19} suffers from large $ n $ divergence. This is an artefact of the infinite size limit of $ \mathcal{N}=2 $ quiver \cite{Lozano:2016kum} that grows unbounded and therefore the space does not close. This is precisely reflected in the range of the $ r $ coordinate which in principle can go to infinity. Therefore, for $ \mathcal{N}=2 $ quivers that are dual to NATD geometries, one must set an upper cut-off on the holographic $ r $- axis for these theories to make sense \cite{Roychowdhury:2023hvq}. It is expected that these large $ r $ divergences will disappear for $ \mathcal{N}=2 $ quivers that are dual to generic Gaiotto-Maldacena (GM) backgrounds, while the quiver is closed by adding flavour degrees of freedom that are sourced due to D6 in the bulk \cite{Lozano:2016kum}.

Using \eqref{e3.20}, the low frequency and low momentum behaviour of \eqref{e3.12} turns out to be 
\begin{align}
\label{e3.21}
\Pi (\tilde{\omega }, \tilde{k})=\frac{\mathcal{D}\tilde{k}^2}{\beta^2 \tilde{k}^4 -(\beta^2 - \alpha^2) \tilde{k}^2 \tilde{\omega}^2 - \alpha^2 \tilde{\omega}^4}
\end{align}
where $ \mathcal{D}=\frac{16}{\pi^2} \log(\frac{n \pi}{4})\frac{R^2}{\alpha'^2}\mathcal{N}_{D4}$ stands for the overall pre-factor.

The coefficients on the other hand, can be expressed as
\begin{align}
\beta^2 =1-\frac{\gamma_E}{\log(\frac{n \pi}{4})} ~;~\alpha^2 = \frac{1}{2 \log(\frac{n \pi}{4})}.
\end{align}

Setting the denominator of \eqref{e3.21} equal to zero yields the pole for the retarded correlators \eqref{e3.11}. The pole yields dispersion relation that exhibits both real and imaginary roots 
\begin{align}
\label{e3.23}
\tilde{\omega}^{(\pm)}_{real}=\pm \tilde{k} ~;~ \tilde{\omega}^{(\pm)}_{img}=\pm \frac{i \beta \tilde{k}}{\alpha}.
\end{align}

The above dispersion relation \eqref{e3.23} represents two classes of ``gapless'' excitations associated with the color sector of $ \mathcal{N}=2 $ quiver. One of these modes move with a constant phase velocity and do not attenuate. The other class of modes, with imaginary dispersion relation, quantum attenuates due to interactions with the color bath. However, none of these modes exhibit imaginary root that goes like $ \sim k^2 $. Therefore, these quasiparticle excitations do not decay like traditional Landau's zero sound modes \cite{Karch:2008fa}-\cite{Pang:2013ypa}. We also notice that these quasiparticles have a different phase velocity as compared to that with the ``first'' sound modes \eqref{e2.14}.

The ground state of $ \mathcal{N}=2 $ quivers are quite exotic as compared to that with standard interacting models with a Fermi surface. This is because the system has both bosonic as well as fermionic excitations. Since color degrees of freedom are strongly correlated in the holographic limit \cite{Lozano:2016kum}, therefore the charged carriers in the color sector interact strongly causing an effect that is similar in spirit to that of the Landau damping \cite{b1}-\cite{b2} in an interacting system with a Fermi surface \cite{Hoyos-Badajoz:2010ckd}. In other words, the low energy (gapless) excitations in the color sector decay through scattering with the color nodes of $ \mathcal{N}=2 $ quiver that acts as a medium.

Finally, we focus our attention on low frequency behaviour of current-current two point function. In the low frequency regime, the two point correlator yields the AC conductivity \cite{Hoyos-Badajoz:2010ckd}
\begin{align}
\label{e3.24}
\sigma (\tilde{\omega})\Big|_{n \gg 1}=-\frac{i}{\tilde{\omega}}\mathcal{G}^{(R)}_{xx}(\tilde{\omega} , \tilde{k} \sim 0)=-\frac{16 i R^2}{\pi^2 \alpha'^2}\mathcal{N}_{D4}\log(n\pi)\tilde{\omega}^{-1}
\end{align}
which goes logarithmically with the rank ($ n \sim N_c $) of the color group in the limit of large $ n $. 

The logarithmic growth is the cumulative effect due to all color nodes pertaining to $ \mathcal{N}=2 $ quiver. The simple ($ \tilde{\omega}=0 $) pole in the AC conductivity \eqref{e3.24} is a typical characteristic feature of conformal fixed points with $ z<2 $ \cite{Hartnoll:2009ns}, where $ z $ is the dynamic critical exponent. The appearance of such a pole is quite suggestive to the fact that the ground state of $ \mathcal{N}=2 $ SCFTs has low energy charge carries. These are precisely the ``gapless'' (quasiparticle) excitations \eqref{e3.23} that are found as poles in the retarded correlator \eqref{e3.21}.  
\section{Correlations at finite density : flavour sector}
\label{flavour}
We now carry out a similar calculation for the flavour sector in $ \mathcal{N}=2 $ quiver, where the flavour carriers are $ U(1) $ charged due to world-volume gauge field $ \mathcal{A}_\mu $ living on $ N_f $ flavour D6 that act as a probe on \eqref{e2.1}. Like before, we consider an expansion $ \mathcal{A}_\mu  \rightarrow \mathcal{A}_\mu +a_\mu (t,r,x) $ of the world-volume $ U(1) $ field, which leads to quadratic action of the following form
\begin{align}
\label{e4.1}
&-S_{D6}^{(2)}=32 \pi N_f T_{D6}\text{Vol}(\mathbb{R}^{1,3})\int_0^{n \pi}dr \frac{L}{\sqrt{\alpha'}}r^2f(r)\Big( \frac{\pi^2  R^2 \alpha'^3 }{L^3}\frac{f^2_{xr}}{\sqrt{(1-\frac{\pi^2 L^4}{R^2}\mathcal{A}_t'^2)(1+\frac{r^2 \alpha'^2}{L^4})}}\nonumber\\
&-\frac{\pi^2 \alpha'^5 }{4L^3}\frac{f^2_{tx}}{\sqrt{(1-\frac{\pi^2 L^4}{R^2}\mathcal{A}_t'^2)(1+\frac{r^2 \alpha'^2}{L^4})}}-\frac{\pi^2  R^2 \alpha'^3 }{L^3}\frac{f^2_{tr}}{\sqrt{(1+\frac{r^2 \alpha'^2}{L^4})}(1-\frac{\pi^2 L^4}{R^2}\mathcal{A}_t'^2)^{3/2}} \Big)
\end{align}
where the fluctuations are turned on along $ t $, $ r $ and $ x $ axes only.

As a next step, we solve these fluctuations in the Fourier space \eqref{e3.2}, with the choice of gauge $ a_r =0 $. The resultant equations of motion, at the linearised level turn out to be
\begin{align}
\label{e4.2}
&\partial_r (h(r)a_x')-\frac{\alpha'^2}{4 R^2}h(r)(\omega^2 a_x + \omega k a_t)=0\\
\label{e4.3}
&\partial_r \Big(\frac{h^3(r)}{r^4}a_t'\Big)+\frac{\alpha'^2}{4 R^2}h(r)(\omega k a_x +  k^2 a_t)=0\\
&k a_x' +\frac{h^2}{r^4}\omega a_t' =0
\label{e4.4}
\end{align}
where, we introduce a new function $ h(r)=\sqrt{r^4+\frac{\pi^2 L^4 C^2}{R^2}} $. Like before, one could show that upon substituting \eqref{e4.4} into \eqref{e4.2}, one precisely reproduces \eqref{e4.3}. In other words, \eqref{e4.2} is redundant and we solve only the last two equations \eqref{e4.3}-\eqref{e4.4}.

Next, we introduce gauge invariant observable $ E(r , \omega , k)=\omega a_x + k a_t $, which yields the following set of solutions
\begin{align}
\label{e4.5}
a'_t=\frac{r^4 k}{r^4 k^2 - h^2(r)\omega^2}E'~;~a'_x = -\frac{h^2(r)\omega}{r^4 k^2 - h^2(r)\omega^2}E'.
\end{align}

Substituting \eqref{e4.5} into \eqref{e4.3}, we find
\begin{align}
\label{e4.6}
E'' + \partial_r \log \Big( \frac{h^3(r)}{r^4 k^2-h^2(r)\omega^2} \Big)E'+\frac{\alpha'^2}{4R^2}\Big( \frac{r^4}{h^2(r)}k^2-\omega^2 \Big)E=0.
\end{align}

The corresponding quadratic action \eqref{e4.1} turns out to be
\begin{align}
-S^{(2)}_{D6}=\mathcal{N}_{D6}\int dr d\omega dk h(r)\Big( E^2-\frac{4R^2}{\alpha'^2}\frac{h^2(r)E'^2 }{(r^4 k^2 - h^2(r)\omega^2)}\Big)
\end{align}
where $ \mathcal{N}_{D6}= 8\pi^3 N_f T_{D6}\alpha'^{7/2}\text{Vol}(\mathbb{R}^{1,3})$ stands for an overall pre-factor.

Using \eqref{e4.6}, one can further simplify the on-shell action as
\begin{align}
\label{e4.8}
S^{(2)}_{D6}\Big|_{\partial r}:=\mathcal{Z}_{f}=\frac{4R^2}{\alpha'^2}\mathcal{N}_{D6}\int d\omega dk \frac{h^3 (r)E'(r)E(r)}{(r^4k^2 -h^2(r) \omega^2)}\Big|^{r= n \pi}_{r= 0}.
\end{align}

Like before, one could show that $ E(r \sim 0)\sim r $, which therefore yields a vanishing contribution to \eqref{e4.8} at $ r=0 $. This naturally satisfies one of the Dirichlet boundary conditions at $ r=0 $. The non-trivial contribution therefore appears due to $ r=n \pi $. Considering $ n $ to be large, one finds the leading order contribution in a large $ n $ expansion as
\begin{align}
\label{e4.9}
\mathcal{Z}_{f}=\frac{4\pi ^2  n^2 R^2}{\alpha'^2}\mathcal{N}_{D6}\int \frac{d\omega dk}{k^2-\omega ^2}E'(r)E(r)\Big|_{r=n \pi}+\mathcal{O}(n^{-2}).
\end{align}

The solution corresponding to \eqref{e4.6} can be obtained in the large $ r(\rightarrow \infty) $ limit, followed by an expansion in the domain of small frequency ($ \tilde{\omega}r\ll1 $) and small momentum ($ \tilde{k}r\ll1 $) such that the ratio $\frac{\tilde{\omega} }{\tilde{k}} $ is finite and much less than unity, where $ \tilde{k}> \tilde{\omega} $. 

Considering a large $r$ expansion of \eqref{e4.6}, one finds an equation
\begin{align}
	E''+\frac{2}{r}E'+\frac{1}{4}(\tilde{k}^2 - \tilde\omega^2)E=0
\end{align}
whose solution can be expressed as\footnote{Similar solution can be obtained by considering an expansion in the reverse order. This follows from the fact that $\frac{\tilde\omega^2}{r^5\tilde{k}^2}\sim 0$ in the limit of small momentum and small frequency, while $r$ is considered to be large enough.}
\begin{align}
\label{e4.11}
E( r, \tilde{\omega}, \tilde{k})=\frac{1}{r}\Big(c_1 e^{-\frac{i r}{2}\sqrt{\tilde{k}^2- \tilde{\omega}^2}}-\frac{i c_2}{\sqrt{\tilde{k}^2- \tilde{\omega}^2}}e^{\frac{i r}{2}\sqrt{\tilde{k}^2- \tilde{\omega}^2}}\Big).
\end{align}
 
 Considering an expansion in the domain of small frequency and momentum, the above solution \eqref{e4.11} yields the two point correlation function in the flavour sector as
\begin{align}
\label{e4.12}
	\Pi(\tilde\omega , \tilde{k})=\frac{2 R^2}{\alpha'^2}\frac{i \tilde{k}^3\mathcal{N}_{D6}}{\tilde{k}^4-\tilde{k}^2 (-|\tilde{C}|^2+\frac{\tilde\omega^2}{2})+\frac{3}{2}\tilde\omega^2 |\tilde{C}|^2}
\end{align}
where we denote $|\tilde{C}|=|\frac{c_2}{c_1}|$ as a new constant of the theory, subjected to the fact $ \frac{|\tilde{C}|^2}{\tilde{k}^2}\ll 1 $. 

One can fix the above constant by demanding the Dirichlet condition, $ E(r \rightarrow \infty)=0 $. Considering the fact that $ r\in [0, n \pi ] $, where $ n $ is a large positive integer, this naturally yields, $ \tilde{C}=-i \nu +\mathcal{O}(\nu^2 r) $ such that, $   \nu r \ll 1 $. Here $ \nu  = \sqrt{\tilde{k}^2- \tilde{\omega}^2} $ is a real positive constant.

Setting the denominator in \eqref{e4.12} equal to zero, one finds a dispersion relation of the form
\begin{align}
\label{e4.13}
	\tilde\omega_{\pm}=\pm  \sqrt{\frac{2}{3}}\tilde{k}\Big(1+  \mathcal{O}(\nu^2/\tilde{k}^2)\Big)
\end{align}
which reveals quasiparticle excitations that move with a constant phase velocity. However, unlike the traditional Landau's zero sound modes, these collective excitations do not exhibit a quantum attenuation through scattering with the medium. This is reflected to the fact that \eqref{e4.13} does not contain an imaginary part that scales like square of the momentum ($\sim \tilde{k}^2$). The other interesting fact to be noticed here is that \eqref{e4.13} represents a different class of quasiparticles, which move with a velocity different from that of the particles in the color sector.

The existence of the ``zero sound'' modes depends on the relative strength between different terms those appearing in the denominator of \eqref{e4.12}. One finds that something similar to zero sound modes appears subjected to the fact that the second term in the expression \eqref{e4.12} can be dropped for certain range of the parameter value $ | \tilde{C} |  \gg \frac{\tilde{\omega}}{\sqrt{2}}+\mathcal{O}(\tilde{\omega}^3/\tilde{k}^2) $. 

This leads to a sub-class of dispersion relation which is of the form
\begin{align}
\label{e4.14}
\tilde{\omega}_{\pm}=\pm i \sqrt{\frac{2}{3 \nu^2}}\tilde{k}^2\Big( 1+ \mathcal{O}(\nu^2/\tilde{k}^2)\Big)
\end{align}
that unlike the other holographic models in the literature, exhibits only a pure imaginary root that goes quadratically with the momentum ($\sim\tilde{k}^2$). In other words, the mode \eqref{e4.14} propagates with a definite ``group velocity'', that quantum attenuates quite similar in spirit as that of the traditional Landau's zero sound modes \cite{Karch:2008fa}-\cite{Pang:2013ypa}. The attenuation \eqref{e4.14} is an artefact of color-flavour interaction in $\mathcal{N}=2$ quiver. To summarise, the low energy spectrum in the flavour sector of $\mathcal{N}=2$ SCFTs contains two types of excitations- (i) one that propagates with a definite (phase) velocity and therefore behaves like a quasiparticle excitation in the usual sense and (ii) the (zero) mode that propagates with a finite group velocity and quantum attenuates through scattering with the ``color'' bath in $\mathcal{N}=2$ SCFTs.

Finally, we estimate AC conductivity, where we set $\tilde{k}=0$ first and thereby take $\tilde{\omega}\rightarrow 0$ limit. Following the above procedure at the level of \eqref{e4.9}, one finds
\begin{align}
	\sigma (\tilde\omega)\Big|_{n \gg 1}=\frac{4\pi i \nu^2 n R^2 \mathcal{N}_{D6}}{\alpha'^2}\tilde\omega^{-3}
\end{align}
which reveals a pole of order three, such that $\nu n \ll 1$ holds true. 

Different behaviour in the pole structure arises due to different nature of the low energy ``collective'' excitations (that are charged under electromagnetic $U(1)$) of the theory. In the previous section, these were low energy ``gapless'' excitations associated with the color sector that are charged under $U(1)$. On the other hand, in the present model, the AC conductivity is sourced due to different type of quasiparticle carriers carrying an electromagnetic $U(1)$ charge, which travel with a different (phase) velocity as compared with the modes in the color sector.
\section{Conclusions and Outlook}
We now summarise the key findings of the paper, along with some future remarks. The purpose of the present paper was to study $\mathcal{N}=2$ SCFTs at finite $U(1)$ charge density and to investigate the low frequency behaviour of the associated two point correlation functions. The pole of the two point function yields the ``spectrum'' of the theory. We also comment about the low frequency behaviour of the AC conductivities by estimating current-current two point function. 

The results that we obtain for the color sector of $ \mathcal{N}=2 $ quivers can be summarised follows. We find no evidence in favor of the holographic ``zero'' sound modes in the color sector of the superconformal quiver. It turns out that the low frequency ``gapless'' excitations of the theory suffer from a ``Landau damping'' through interactions with other degrees of freedom in the color multiplet of the quiver. This seems to be a reasonable model since the color nodes of the quiver are strongly interacting in the holographic limit \cite{Lozano:2016kum}. The AC conductivity, on the other hand, exhibits a simple pole in the zero frequency limit, which we identify as the characteristic feature of $ z=1 $ conformal fixed point \cite{Hartnoll:2009ns}, \cite{Hoyos:2011zz}. This also signifies that the low energy excitations of the theory are indeed the massless carriers carrying a $ U(1) $ charge.

When flavours are added as a probe, we obtain a different dispersion relation altogether. It turns out that unlike in the color sector, the low energy spectrum contains propagating modes that travel with a group velocity and quantum attenuates quite similar as that of Landau's zero sound.  This attenuation is sourced due to interactions between the color and the flavour modes of $\mathcal{N}=2$ quiver. The dispersion relation also contains quasiparticle excitations that propagate with a (phase) velocity which is different from what has been observed for the color sector in $\mathcal{N}=2$ quiver. This also gets reflected in the AC conductivity, which exhibits a cubic pole structure at low frequencies, different from what has been observed for the color sector. Our results show that $\mathcal{N}=2$ quivers fall under the same universality class along with other canonical QFTs with dynamic critical exponent $z=1$ \cite{Karch:2008fa} as well as $1\leq z <2$ \cite{Hoyos-Badajoz:2010ckd}. In all these theories we came across zero sound modes that behave like quasiparticle excitation above some ground state. However, such a zero mode ceases to exist for theories with $z\geq 2$ \cite{Hoyos-Badajoz:2010ckd}.

It would be really nice to repeat all these calculations for generic Gaiotto-Maldacena (GM) class of geometries that are dual to $\mathcal{N}=2$ quivers of finite size. These quivers (as compared to the quivers studied in the present paper) are complete in the sense that there are flavour nodes attached at the end points of the long color chain. This would require to adopt to a new set of coordinates, where the metric functions can be expressed in terms of a potential function satisfying Laplace's equation \cite{Lozano:2016kum}. This set up would require a careful choice of embedding for probe D- branes as the geometry involves singularities near the location of the flavour D6. Finally, it would be also interesting to repeat above calculations for $\mathcal{N}=1$ SCFTs in 5d \cite{DHoker:2016ujz} or $\mathcal{N}=(0,1)$ SCFTs in 6d \cite{Apruzzi:2013yva}, whose holographic descriptions are well studied in the literature.
\subsection*{Acknowledgments}
The author is indebted to the authorities of IIT Roorkee for their unconditional support towards researches in basic sciences. The author would also like to acknowledge The Royal Society, UK for financial assistance. Finally, the author acknowledges the Mathematical Research Impact Centric Support (MATRICS) grant (MTR/2023/000005) received from ANRF, India.


\end{document}